\documentclass{aa}  
\usepackage{natbib}
\bibpunct{(}{)}{;}{a}{}{,}

\usepackage{graphicx}
\usepackage{hyperref}
\usepackage{xcolor}

\usepackage{txfonts}

\newcommand{\hip}{HIP~67522\;}
\newcommand{\radioflux}{$\text{erg s}^{-1}\text{Hz}^{-1}$}
\newcommand{\lum}{$\text{erg s}^{-1}$}

\begin{document}

   \title{Searching for planet-induced radio signal from the young close-in planet host star HIP 67522}

  \author{E. Ilin
          \inst{1}
          \and
          S. Bloot\inst{1,2}
          \and
          J. R. Callingham \inst{1,3}
          \and
          H. K. Vedantham \inst{1,2}
          }

   \institute{ASTRON, Netherlands Institute for Radio Astronomy, Oude Hoogeveensedijk 4, Dwingeloo, 7991 PD, The Netherlands\\
              \email{ilin@astron.nl}
         \and
             Kapteyn Astronomical Institute, University of Groningen, P.O. Box 800, 9700 AV, Groningen, The Netherlands
         \and
            Anton Pannenkoek Institute for Astronomy, University of Amsterdam, Science Park 904, 1098\,XH, Amsterdam, The Netherlands
             }

   \date{Received Mar 21, 2025; accepted June 3, 2025}

  \abstract{
   \hip is a 17 Myr old 1.2\,$M_\odot$ dwarf, and the only such young star known to host two close-in gas giant planets. The inner planet likely orbits close enough to its host to power magnetic star-planet interactions. In the radio domain, magnetic star-planet interaction is expected to produce a unique signature: electron cyclotron maser emission (ECME), beamed in phase with the orbit of the close-in planet.
   We conducted the longest radio monitoring campaign of a G dwarf host star to date to search for ECME, totaling $135\,$h on \hip over a period of five months with the Australia Telescope Compact Array (ATCA) between $1.1-3.1$\,GHz.
   
   We find that \hip is active in the radio, with emission that varies stochastically, with a duty cycle of $69\%$ above $0.24\,$mJy, and frequent bursts. Both the bursts and the quiescent emission are consistent with the canonical G\"udel-Benz relation, and show a positive spectral index and brightness temperatures $\geq 10^{10}\,$K, indicating likely a flaring origin. Our observations cover $61\%$ of the innermost planet's orbit, including multiple visits of the quadrature where planet-induced ECME detection is most likely for this system. However, no orbital modulation or persistent polarization of the radio emission was detected. Our upper limit on circularly polarized emission from \hip suggests a $<0.7\%$ conversion efficiency of the magnetic power generated in the star-planet interaction into radio waves, unless the emission was missed by our phase or frequency coverage, or was absorbed in the circumstellar plasma. \hip is a system with one of the highest expected powers of star-planet interaction among known systems and shows strong indication of planet-induced flaring, motivating observations at other wavelengths to probe for further dissipation pathways.
   }


   \maketitle

\section{Introduction}

Recently, many close-in planets have been found to experience harsh space weather conditions. Their atmospheres react to the bombardment with their host star's X-ray and UV emission by heating up and expanding, which can be measured as low bulk or atmospheric density~\citep{laughlin2011anomalous, barkaoui2024extended, thao2024featherweight}, or spectroscopic signatures of escaping material~\citep{lecavelierdesetangs2010evaporation, ehrenreich2015giant, spake2021posttransit}. After millions to billions of years of atmospheric erosion, some planets end up with a much smaller atmosphere than they were born with, and others with no atmosphere at all~\citep{tian2015atmospheric, ketzer2023influence, vanlooveren2024airy}. 
On the flip side, a sufficiently close-in planet may also influence its host star's evolution~\citep{shkolnik2018signatures}. The planet presents an obstacle to the star's magnetic field, and may set off Alfv\'en waves in the stellar magnetosphere that can propagate back to the star. A detectable fraction of the energy carried by the Alfv\'en waves could be emitted in the radio wave band via the electron cyclotron maser emission (ECME) in the radio~\citep{zarka2001magneticallydriven,treumann2006electroncyclotron,hess2011modeling,kavanagh2021planetinduced,callingham2024radio}.

Searches for planet-induced ECME have so far been inconclusive. The $\tau$ Boo system hosts a close-in Hot Jupiter, and has shown tentative low-frequency ECME~\citep{turner2021search}, but the signal could not be recovered in follow-up observations~\citep{turner2024followup,cordun2025deep}. ECME has also been detected on GJ 1511, a star perhaps too inactive to produce ECME purely through stellar activity~\citep{vedantham2020coherent}, but a planet in close orbit has yet to be detected~\citep{blanco-pozo2023carmenes}. Similarly, no planets have been detected in close orbits around brown dwarfs yet. Meanwhile, ECME is the main source of their radio emission~\citep{hallinan2008confirmation, hallinan2015magnetospherically}. Close-in planets have been detected around AU Mic~\citep{plavchan2020planet, martioli2021new}, and also ECME~\cite{bloot2024phenomenology}, but the emission was modulated with the rotational period of the star, and could not be linked to the presence of planets directly. Hints of ECME modulated with a planet's orbital period have been suggested for YZ~Cet~\citep{pineda2023coherent} and Proxima~Cen~\citep{perez-torres2021monitoring}. However, long-term monitoring is required to rule out coincidental intrinsic stellar emission.

Long-term monitoring for ECME requires careful target selection. The \hip system~(Table~\ref{tab:params}) is a 17 Myr member of the Upper Centaurus Lupus part of the Sco-Cen OB association~\citep{ dezeeuw1999hipparcos,rizzuto2020tess} at a distance of $125\,$pc~\citep{gaiacollaboration2021gaiaa}. The host star is a $1.2M_\odot$ dwarf~\citep{rizzuto2020tess} with a low-density giant planet with an expanded atmosphere~\citep{thao2024featherweight} in a $6.95\,$d orbit~\citep{rizzuto2020tess}, and another recently confirmed gas giant in a $14.33\,$d orbit~\citep{barber2024tess}. The star is oriented equator-on and the innermost planet has low obliquity~\citep{heitzmann2021obliquity}. With one of the highest predicted powers produced in star-planet interactions ~\citep{strugarek2022moves,ilin2024planetary}, and a recent detection of significant clustering of flares with orbital phase that strongly suggest planet-induced emission~\citep{ilin2025closein}, \hip is a promising target for a search for periodic bursts of interaction-driven ECME, unambiguously identified by its occurrence in phase with the orbit of the interacting planet. 

\begin{table}[]
    \caption{Stellar parameters of \hip\hspace{-0.1cm}, and orbital parameters of \hip b.}
    \begin{tabular}{l c}\hline\hline
         & HIP~67522 \\\hline
      TIC ID & 166527623 \\   
      RA (J2000) [hh:mm:ss] & $13~50~06.28$\tablefootmark{1}\\
      Dec (J2000) [dd:mm:ss] & $-40~50~08.88$ \tablefootmark{1}\\
      $d$ [pc]   & $124.7 \pm 0.3$ \tablefootmark{1} \\
      age [Myr] & $15-20$ \tablefootmark{2,3} \\
      $T_{\rm eff}$ [K] & $5650\pm75$\tablefootmark{3} \\
      $M$ [$M_\odot$] & $1.2 \pm 0.05$\tablefootmark{3} \\
      $R_*$ [$R_\odot$] & $1.39 \pm 0.06$\tablefootmark{3} \\\hline
      & HIP 67522 b\\\hline
      $P_{\rm orb}$ [d] & $6.959473 \pm 0.000002 $ \tablefootmark{4} \\
      $T_0$ [JD] & $2458604.02376 \pm 0.0003 $ \tablefootmark{4}\\\hline
    \end{tabular}
    \label{tab:params}
\tablefoot{
\tablefoottext{1}{\citet{gaiacollaboration2021gaiaa}}
\tablefoottext{2}{\citet{dezeeuw1999hipparcos}}
\tablefoottext{3}{\citet{rizzuto2020tess}}
\tablefoottext{4}{\citet{barber2024tess}}
}
  
\end{table}

\section{Radio observations and analysis}
\label{sec:data}

We observed HIP~67522 from April 7 to July 15, 2024, with the Australian Telescope Compact Array (Project ID: C3591, PI: Ilin) for a total of $134.7\,$h on target, searching for periodic occurrence of ECME~(Fig.~\ref{fig:radio_grid} and Table~\ref{tab:atca_obs_log}). We conducted observations in L/S band, i.e. $1.1-3.1\,$GHz, with a $10$\,s integration time and $1\,$MHz channel width using the $2\,$GHz Compact Array Broadband Backend~\citep[CABB,][]{wilson2011australia}.

We observed PKS~B0823-500 as one of our primary calibrators at the start of the observations, because the standard primary calibrator for L band, PKS~B1934-638, was usually at too low an elevation at the start of our runs. We also observed PKS~B1934-638 at the end of each run. PKS~B0823-500 and PKS~B1934-638 were each observed for 10\,min. During the rest of each observation, we alternated between 3 min on the secondary calibrator, PKS~B1424-418, and 40 min on target.

We reduced the observations with CASA (version 6.6.0.20, \citealt{team2022casa}). Radio frequency interference (RFI) in the $1.1-3.1\,$GHz band was ubiquitous, which led to flagging of $30-60\%$ of the data before calibration. We used the \texttt{TFCrop} algorithm that identifies outliers in the time-frequency plane, cutting at $3.5$ and $4$ standard deviations in frequency and time, respectively. The baseline for the cut is calculated using a polynomial and linear fit in frequency and time, respectively. After visual inspection, further manual flagging was required in some cases.

We used PKS~B1934-638 as the flux density and bandpass calibrator for all except one observation. On June 26, 2024, the observation of PKS~B1934-638 was subject to so much RFI that the calibration solutions did not converge, so we used PKS~B0823-500 instead. For phase calibration, we used PKS~B1424-418 in all runs.

We calculated the complex gains, spectral bandpass, polarization and polarization leakage solutions from PKS~B1934-638 or PKS~B0823-500, using CA06 as reference antenna for the phase solutions, and $60\,$s as solution interval, during which we assumed that the phase remained constant. After a first phase calibration on the bandpass calibrator, we performed a bandpass calibration on the same object, which were applied to the primary (bandpass) and secondary calibrator data. We transferred the solutions for the bandpass and gain to derive the polarization calibration using the bandpass calibrator and the \texttt{qufromgain} task from \texttt{casarecipes.atcapolhelpers}. 

As the final product, we made images of the target based on the calibrated measurement sets, and extracted point source flux density and background from each image. We used \texttt{tclean} with a manual mask to construct images in different time intervals, including entire observing runs of about $5-10\,$h each, $1\,$h for the general time series, and $30\,$min during a large burst. In all observations, we removed antenna CA03, which suffered from an instrumental defect throughout our campaign. In Stokes I, we also removed baselines shorter than $500\,$m (in the H168 configuration, this removes all but 4 baselines) to further reduce RFI and contamination from the side lobes of a bright resolved source about 9' away from the target. Due to the bright source, we used the \texttt{multiscale} algorithm for the minor cycles of the cleaning process in Stokes I, and continued the cleaning until the target emission was reduced to background levels. In Stokes V, no other sources appeared nearby our target, so we used the \texttt{hogbom} point source model instead. We used \texttt{imstat} to extract the point source flux density as the maximum value in the beam of the target, and background as the standard deviation of a nearby, source-free region from the final images. 

We detected one large burst, of which we extracted the dynamic spectrum from the observations in Stokes I, following the approach outlined by~\citet{bloot2024phenomenology}. We imaged the observations in $6000\times6000$ pixels, i.e., 2.4''$\times$2.4'' pixel size, using WSClean~\citep{offringa2014wsclean} to generate a model of all sources but HIP 67522, subtracted all sources but \hip from the image with the CASA task \texttt{uvsub}, then phase-shifted the measurements at each epoch with \texttt{DP3}~\citep{vandiepen2018dppp} to the location of \hip, which we determined from the maximum flux pixel in the residual image. 

\begin{table*}
\centering
\caption{ATCA observing log.}             
\label{tab:atca_obs_log}      
\begin{tabular}{llllll}
\hline\hline
Obs. Date & Obs. Start & Obs. Stop & Duration\tablefootmark{a} & Stokes I & Array \\

[UTC] & [UTC] & [UTC] & [h] & [mJy] &  Config. \\ \hline
2024-04-07 & 09:10 & 20:20 & 10.2 & <0.10 & 6A \\
2024-04-21 & 08:39 & 16:42 & 7.4 & 0.19 [0.03] & 6A \\
2024-04-26 & 08:46 & 19:36 & 9.9 & 0.24 [0.02] & 6A \\
2024-04-28 & 13:04 & 18:10 & 4.7 & <0.21 & 6A \\
2024-05-05 & 13:14 & 18:40 & 5.0 & 0.27 [0.03] & 6A \\
2024-05-11 & 07:00 & 18:15 & 10.3 & 0.38 [0.02] & 6A \\
2024-05-15 & 06:37 & 17:54 & 10.3 & 0.21 [0.02] & 6A \\
2024-05-31 & 05:24 & 16:35 & 10.2 & <0.07 & H168 \\
2024-06-11 & 06:10 & 12:14 & 5.5 & 0.98 [0.02] & 6D \\
2024-06-14 & 04:28 & 15:49 & 10.4 & 0.33 [0.02] & 6D \\
2024-06-19 & 05:00 & 15:44 & 9.1 & 0.29 [0.02] & 6D \\
2024-06-26 & 03:49 & 15:13 & 10.4 & 0.36 [0.01] & 6D \\
2024-06-26 & 03:49 & 15:13 & 10.4 & 0.36 [0.01] & 6D \\
2024-07-04 & 02:23 & 14:04 & 10.7 & 0.19 [0.02] & 6D \\
2024-07-15 & 01:44 & 13:13 & 10.2 & <0.07 & 6D \\
 &  &  & $\Sigma = $134.7 &  &  \\
\hline
\end{tabular}
\tablefoot{
\tablefoottext{a}{The duration marks the total time on target, excluding the periodic scans of the secondary calibrator.}}
\end{table*}

We defined any observation as a detection if the signal-to-noise ratio was $S/N>4$, and coincided with a point source in the image at the location of \hip\hspace{-.1cm}~(see Fig.~\ref{fig:Appendix}). Otherwise, we defined the upper limit as four times the RMS noise.
We computed the radio luminosity, $L_R$ and the characteristic brightness temperature $T_b$, assuming that the entire stellar disk at $1R_*$ is a uniformly emitting area, that the emission is isotropic, and using a distance $d=124.7\,$pc~(see Table~\ref{tab:params}),

\begin{equation}
    T_b = \dfrac{d^2 c^2 S_\nu}{R_*^2 2 \pi \nu_0^2 k_B},
\end{equation}

where $c$ is the speed of light, $\nu_0$ the center frequency in L band at $2.1\,$GHz, $S_\nu$ the flux density, $R_*$ and $d$ the photometric radius and distance of the star, and $k_B$ the Boltzmann constant. For the only observing run with a Stokes V detection, on April 26, 2024, we additionally calculated the degree of circular polarization as the ratio between the average flux density in Stokes V and Stokes I.

\section{Results}
\label{sec:results}

  \begin{figure*}
  \includegraphics[width=0.92\hsize]{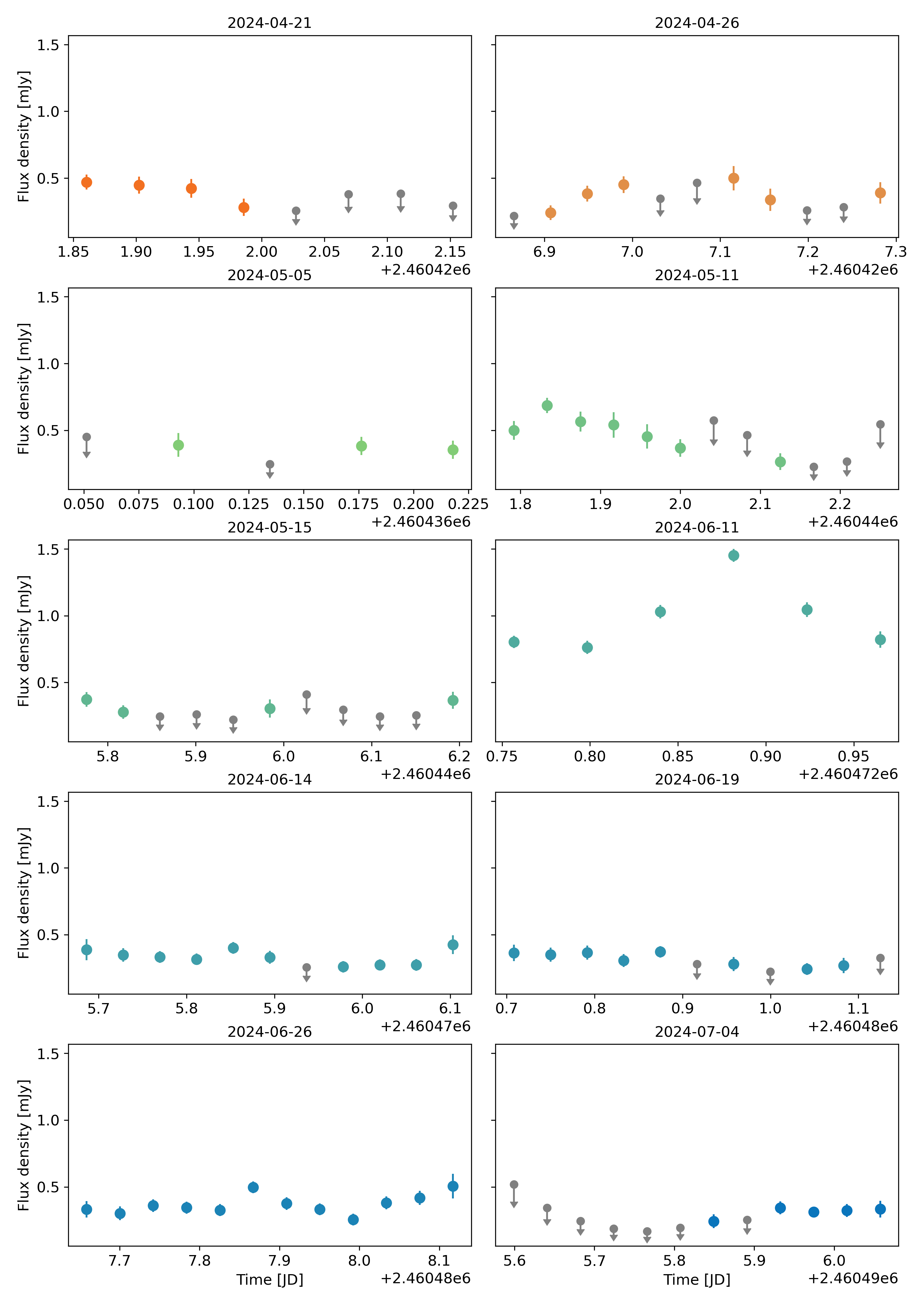}
      \caption{Light curves of Stokes I flux density in the $1.1-3.1\,$GHz band of ATCA, with one panel for each observing run where \hip was detected. Gray arrows indicate non-detections, with arrow length corresponding to the $1-\sigma$ RMS noise.}
         \label{fig:radio_grid}
   \end{figure*}

   \begin{figure}
   \centering
   \includegraphics[width=\hsize]{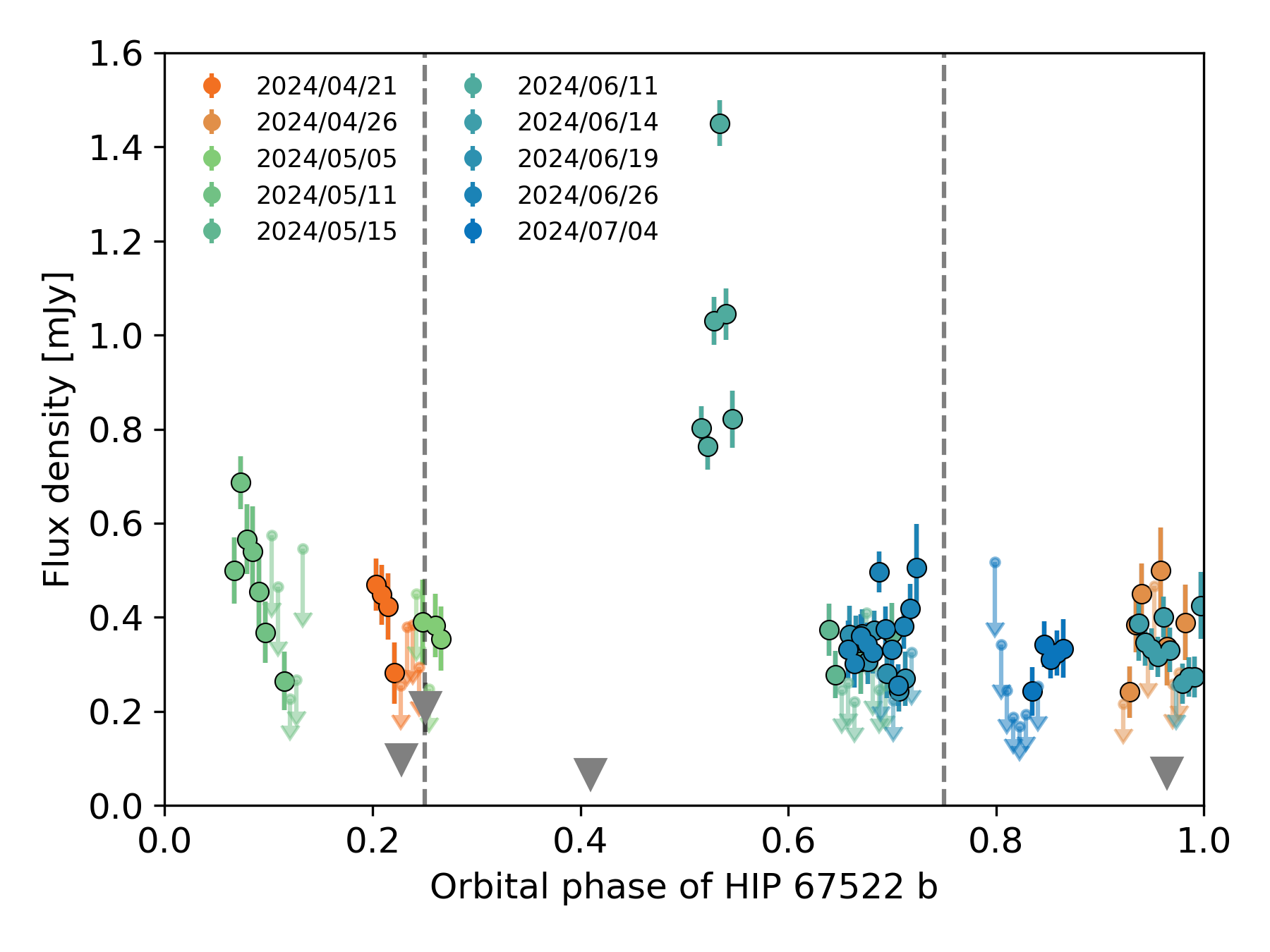}
      \caption{ATCA Stokes I light curves from Fig.~\ref{fig:radio_grid} folded with the orbital period of \hip\,b (transit midtime set to phase 0). Colored downward arrows indicate non-detections in a 1-h interval of a run with a significant source detection in the integrated image, gray triangles indicate that the entire observing run yielded a non-detection. Dashed gray lines mark the quadratures of HIP 67522\,b. The Stokes I emission does not show any variability in phase with the planet's orbit. Error bars indicate $1-\sigma$ RMS noise.
              }
         \label{fig:results:phasedradio}
   \end{figure}

The quiescent flux density in the $1.1-3.1\,$GHz band of ATCA was on average $0.26 \pm 0.02\,$mJy, or a luminosity of $(4.8\pm 0.4)\times 10^{15}\,$\radioflux~obtained from the full integration images of each run with detections, excluding the June 11 and May 11 observations, which we treated as burst episodes. The corresponding brightness temperature at the central wavelength of $2.1\,$GHz was $(1.0\pm0.1) \times 10^{10}\,$K. The radio emission was unpolarized in all observations except for the run on April 26, where we marginally detect a polarization fraction $S_V/S_I=22\pm6\,\%$ for the full run. However, we could not temporally resolve the polarization signal due to a low signal-to-noise ratio of the emission.

The quiescent emission was not detectable in all ATCA epochs, which was sometimes due to worse RFI conditions that increased the noise of an observation, but also due to lower emission from the star. At 1-h time resolution, the lowest detected flux density was in the April 21 epoch at \mbox{$F_{\rm thresh} = 0.24\pm0.07\,$mJy}. We defined the emission in \hip as "on" for any flux density $F \geq F_{\rm thresh}$, and as "off" whenever no flux was detected and the upper limit was below $F_{\rm thresh}$. With this definition, we found a duty cycle of radio emission of $\sim69\%$. Folding the light curves with the orbital period of \hip\,b does not reveal any significant modulation in phase with the planet's orbit~(Fig.~\ref{fig:results:phasedradio}). The polarized emission on April 26 occurred close to transit, but was not detected during any of the two later observations in that phase range.

\begin{figure}
   \centering
   \includegraphics[width=\hsize]{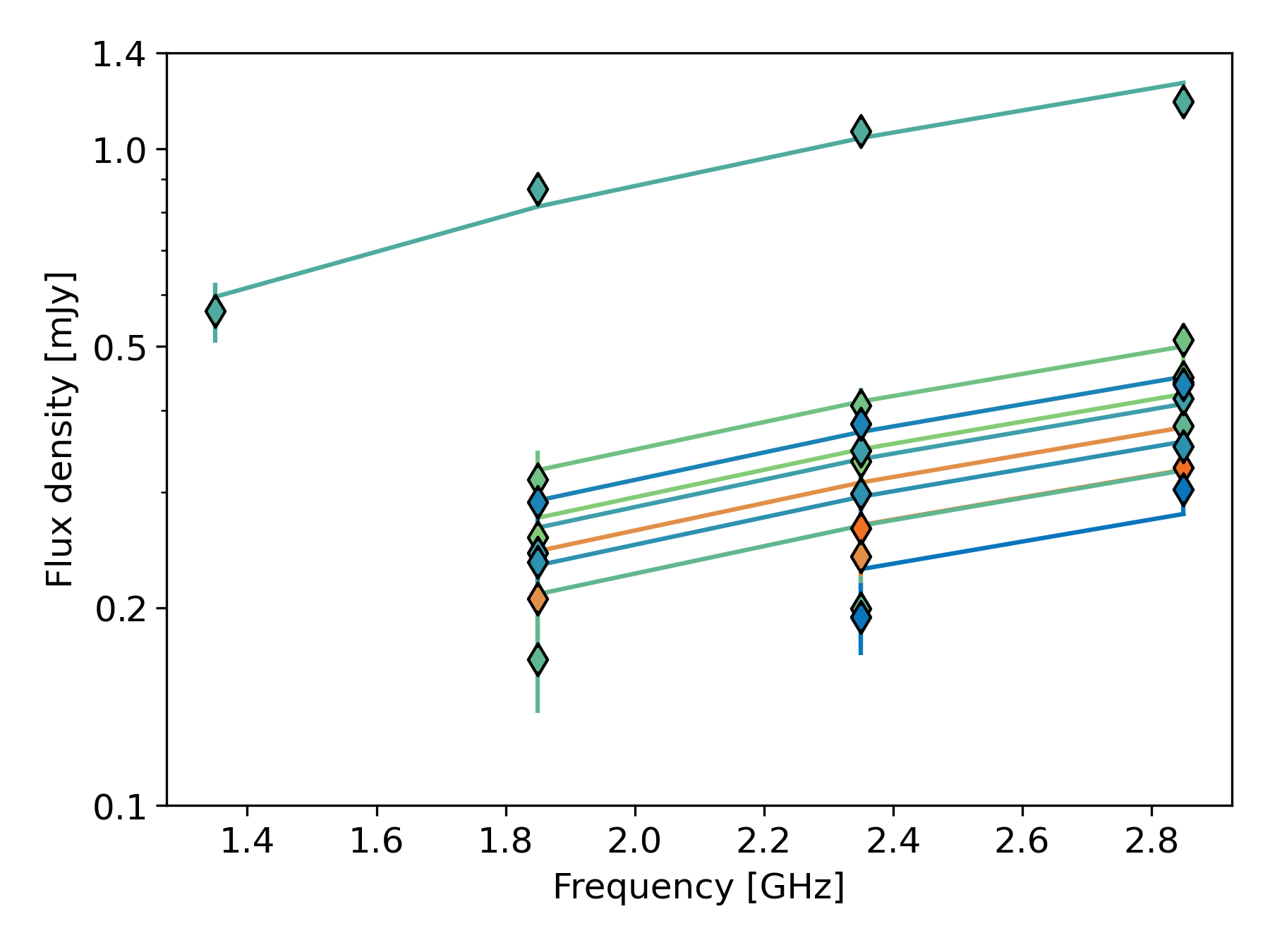}
      \caption{Radio spectra of \hip (diamonds, same color-coding as in Fig.~\ref{fig:results:phasedradio}), and power law fits (lines). Fitting all spectra jointly, as shown here, yields a spectral index of $\approx 1.0$. }
         \label{fig:results:atcaspectra}
   \end{figure}

In the epochs with detected emission, we split the frequency band in four $500\,$MHz-wide sub-bands, and extracted images following the procedure described in Section~\ref{sec:data} for the full runs. From the four bands we then calculated the spectral index assuming $S_\nu\propto\nu^\alpha$. Fitting spectra from individual runs yielded spectral indices ranging from $0.8$ to $2.1$, but the higher indices were only associated with the faintest three spectra. Assuming that all spectra share the same index, we performed a joint least square fit to the linearized power law, which yielded $\alpha\approx 1.0$, suggesting that the emission was optically thick at $1.1-3.1\,$GHz~(Fig.~\ref{fig:results:atcaspectra}). 

\begin{figure}
\centering
\includegraphics[width=\hsize]{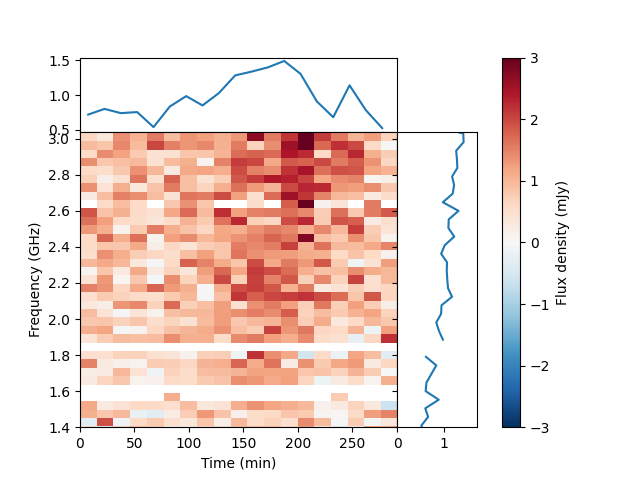}
  \caption{Dynamic spectrum of the June 11, 2024, radio burst. We omitted frequencies below $1.4\,$GHz, which were dominated by RFI. The burst mostly occupies higher frequencies, likely beyond the $3.1\,$GHz limit, and its spectrum shows a positive slope~(right panel, see also Fig.~\ref{fig:results:atcaspectra}). 
          }
     \label{fig:results:dynspec}
\end{figure}

Over the course of $134.7\,$h, \hip showed one strong burst on June 11~(Fig.~\ref{fig:results:dynspec}), and a potential burst seen as a turnover at the beginning of the observing run on May 11, followed by a prolonged emission decay~(see second row, right column in Fig.~\ref{fig:radio_grid}). The flux density rose to peak values of $1.45\,$mJy and $0.69\,$mJy, or \mbox{$2.7\times10^{16}\,$\radioflux} and \mbox{$1.3\times10^{16}\,$\radioflux} on June~11 and May~11, respectively. The corresponding brightness temperatures are $5.4\times10^{10}\,$K and $2.6\times10^{10}\,$K, respectively. The June~11 burst was bright enough to extract a dynamic spectrum from the data~(Fig.~\ref{fig:results:dynspec}). The spectrum shows a positive spectral index consistent with the quiescent emission~(Fig.~\ref{fig:results:atcaspectra}), but no obvious drift in frequency over time.

\section{Discussion}

The quiescent radio emission from \hip is typical of coronal emission in other cool dwarfs~(Section~\ref{sec:discussion:gbr}), and the properties of both bursts and quiescence are compatible with emission from accelerated electrons during flares~(Section \ref{sec:discussion:radiomechanism}). Assuming that flares occurring at the same rate in both optical and radio correspond to events of the same total energy, we can use the optical flare rate to infer the flare energies corresponding to the detected bursts. We find that neither of the bursts could be detected with Transiting Exoplanet Survey Satellite~\citep[TESS,][]{ricker2015transiting}, and only the brightest could be marginally detected with the CHaracterising ExOPlanet Satellite~\citep[CHEOPS,][]{benz2021cheops}. The radio emission does not show any signs of star-planet interaction, which may indicate a low efficiency of dissipation via ECME, but may also be explained by reabsorption of the emission, beaming effects, or observational constraints~(Section~\ref{sec:discussion:spi}).

\subsection{G\"udel-Benz relation}
\label{sec:discussion:gbr}

The G\"udel-Benz relation (GBR) identifies coronal emission by the relation between centimeter radio and X-ray luminosity, originally derived from ROSAT~\citep{truemper1982rosat} observations at $0.1-2.5\,$keV, and Karl G. Jansky Very Large Array\citep[VLA,][]{perley2011expanded} observations at $5-8.5\,$GHz~\citep{guedel1993xray, gudel1994discovery}. The relationship can also help with determining incoherent and coherent radio emission processes \citep{callingham2024radio}. We use the fit derived by~\citet{williams2014trends}, who extended the observations to ultracool dwarfs with Chandra ($0.2-2.0\,$keV) and VLA in \citet{berger2010simultaneous}:

\begin{equation}
    \log_{10} L_R = 1.36 \times (\log_{10} L_X - 18.97)
\end{equation}
\citet{maggio2024xuv} report a quiescent X-ray luminosity for \hip~in XMM-Newton's $0.3-10\,$keV band of
$\log_{10} L_X = 30.5\,$\lum. For comparison with the above relation, we computed the X-ray luminosity of \hip in the $0.2-2.0\,$keV band using the quiescent emission parameters in Table~1 in~\citet{maggio2024xuv} using \texttt{pyxspec}~\citep{gordon2021pyxspec}, which yielded \mbox{$\log_{10} L_X = 30.8\,$\lum}, and \mbox{$\log_{10} L_R = 16.0\pm 0.6\,$\radioflux}, predicted by GBR. The mean quiescent flux in ATCA's L/S band observations was $\log_{10} L_R = 15.69\pm0.04\,$\radioflux. We can use the spectral index of $\approx 1$ and the mean quiescent flux density from our observations to extrapolate what $L_R$ would be at the mid-frequency of the VLA band, to which the GBR is calibrated, which yields \mbox{$\log_{10} L_R=16.2\pm 0.4\,$\radioflux}, consistent with the GBR prediction. The high radio luminosity of \hip puts it in company with other young stars, e.g., T Tauri and RS CVn~\citep{guedel1993xray,yiu2024radio,davis2024detection}. The consistency of the emission with the GBR suggests most likely the radio emission is incoherent in nature.

\subsection{Radio emission mechanism}
\label{sec:discussion:radiomechanism}
\hip\hspace{-.1cm}'s emission is characterized by a brightness temperature $\geq10^{10}\,$K, a positive spectral index $\alpha\approx 1$, burst-like variability, weakly polarized emission, and consistency with the GBR. The high brightness temperature rules out thermal emission, which is limited to $\leq10^8\,$K for a typical corona on a solar type star~\citep{gudel2004xray}. The detected bursts, and the consistency with the GBR suggest that the emission is related to particle acceleration in flares~(see~\citealt{yiu2024radio,davis2024detection} for similar conclusions). \citet{osten2005radio} measured positive spectral indices at $5\,$GHz and low degree of polarization during the early phases of a large flare on the active M dwarf EV~Lac. As in their work, our spectral index is too flat to be consistent with typical power law electron distributions of the form $n(E)\propto E^{-\delta}$ with $\delta \approx 2-7$~\citep{dulk1982simplified,dulk1985radio}, potentially indicating multiple co-existing electron populations. In~\citet{osten2005radio}, the late decay phases showed a moderate degree of circular polarization and negative spectral index, indicating a decrease in optical depth, which we did not detect at any time during our monitoring. We speculate that the L/S band emission in \hip was dominated by the combined peak emission of a population of frequent flares, two of which were bright enough to pass the detection threshold as individual bursts. 

If the June 11 and May 11 bursts were directly related to flares, we can use the full observing duration of $\Delta t = 134.7\,$h and compare the occurrence rate to the corresponding flare rate in the optical flare frequency distribution, derived from flares observed with TESS~\citep{ilin2024planetary} and CHEOPS~\citep{ilin2025closein}. The flux turnover in the May 11 burst overlaps with a visit by CHEOPS, but no optical flare was detected during or prior to the burst. Yet, the non-detection is consistent with the burst being associated with an undetected flare. The energy of a flare that would occur twice per $\Delta t$ would have a bolometric energy $5\times10^{33}\,$erg, below the detection threshold of TESS and marginally detectable with CHEOPS, assuming rate equivalence as discussed above. However, the instrumental noise level during that particular CHEOPS visit was higher than average, so that a flare of that energy could remain undetected.

\subsection{Star-planet interaction}
\label{sec:discussion:spi}
In the radio band, we did not find any highly circularly polarized emission that could identify a magnetic interaction between the planet and the star, or any other modulation in radio emission with the orbital phase of \hip\,b~(Fig.~\ref{fig:results:phasedradio}). Assuming a simple stellar dipole magnetic field aligned with the rotation axis of the star, and an opening angle of the emission cone of EMCE of nearly $90\,$deg, planet-induced emission should be detected around quadrature phase, which was visited four times during the campaign. A more complex field could produce ECME that may have been missed by our $61\%$ phase coverage, or beam the emission away from the line of sight entirely. But even if the dipole field assumption is correct, an absence of periodically recurring electron cyclotron maser emission (ECME) does not rule out the presence of star-planet interaction for three reasons. First, ECME may be present, but become reabsorbed during propagation in the stellar magnetosphere~\citep{melrose1982electroncyclotron}. Second, the conditions for ECME to escape may be present but only further out in the magnetosphere, i.e. at weaker magnetic fields strengths. The emission wavelength $\nu_B = 2.8 \times B\,$[G] MHz and its low order harmonics would then be seen at frequencies observable by instruments like LOFAR, i.e., 144\,MHz~\citep{pope2021tess, callingham2021population}, and not in ATCA's gigahertz bands. Third, multiple mechanisms~\citep{lanza2018closeby,saur2013magnetic} are potentially capable of dissipating the energy from star-planet interactions in the stellar corona or chromosphere~\citep{scharf2010possible, poppenhaeger2011correlation,shkolnik2008nature,cauley2019magnetic}. Dissipation via the ECM channel requires particular conditions that convert the Alfv\'en wave energy into an unstable particle distribution. 

If some of the interaction flux dissipates and escapes through the ECME mechanism in ATCA's L/S~band, we can place an upper limit based on our non-detections of the system in Stokes V. The average upper limit throughout the observing campaign, calculated as the four-fold RMS noise, was $1.1\,$mJy, based on a $30\,$min long burst spread over the full integration, which would correspond to an emission cone width of about 1\,deg. Assuming an average surface field strength of $2100\,$G~\citep{ilin2024planetary} and that the large-scale field strength is $\sim 20\%$ of that value for very active stars~\citep{reiners2022magnetism,kochukhov2020hidden}, we can rule out ECM radio power values in excess of $4.3\times 10^{23}\,$erg/s. The corresponding efficiency is approximately $7\times 10^{-3}$, based on the estimated power of interaction of $6\times 10^{25}\,$erg/s in the Alfv\'en wing mechanism~\citep{saur2013magnetic}. This upper limit is consistent with the conversion rate of $\sim 2-10\times 10^{-3}$ of magnetic interaction power to radio emission between Jupiter and its Galilean moons~\citep{zarka2007plasma}. Therefore, the non detection of Stokes V emission does not rule out magnetic star-planet interaction in the Alfv\'en wing scenario.

\section{Summary and conclusions}
In our 134.7\,h radio observing campaign, \hip appears active and variable, including two bursts, as expected for its young age. The high brightness temperature of the emission, its adherence to the G\"udel-Benz relation~(see also~\citealt{maggio2024xuv}), the low degree of polarization, and the positive spectral index suggest that the radio emission is non-thermal and of coronal origin, likely stemming from particle acceleration during flares.

The radio emission showed no modulation in phase with the orbital period of the planet. The upper limit on Stokes V emission from our observations suggests that Alfv\'en waves generated by star-planet interaction dissipate less than $0.7\%$ of their power via ECME. However, the absence of measured emission can also be explained by reabsorption in the stellar plasma environment, or by emission at frequencies not captured by ATCA's $1.1-3.1\,$GHz band. Alternatively, the emission may have occurred at orbital phases not covered by our campaign, or beamed away from the line of sight entirely.  

Our results place first constraints on the flux and duty cycle of ECME in a star-planet system with strong clustering of flares in planetary orbital phase that indicates a planet-induced origin~\citep{ilin2025closein}. Future (multiwavelength) observations probing the stellar large scale magnetic field and various dissipation pathways of planet-induced perturbations will be instrumental in deciphering the geometry and energetics of magnetic star-planet interaction in this system. In the radio domain, extended frequency coverage (e.g., with SKA Low,~\citealt{dewdney2009square}) and an order of magnitude increase in total observing time would allow us to measure planet-induced modulation of unpolarized radio emission from flares triggered by magnetic interaction~\citep{ilin2022searching,ilin2024planetary, fischer2019timevariable}. 

\begin{acknowledgements}
The authors thank the anonymouse referee for helpful comments that improved the manuscript.

The Australia Telescope Compact Array is part of the Australia Telescope National Facility, which is funded by the Australian Government for operation as a National Facility managed by CSIRO. We acknowledge the Gomeroi people as the traditional owners of the Observatory site. 

This paper includes data collected by the TESS mission, which are publicly available from the Mikulski Archive for Space Telescopes (MAST).

E.I. and H.K.V. acknowledge funding from the European Research Council under the European Union's Horizon Europe program (grant number 101042416 STORMCHASER). S.B. acknowledges funding from the Dutch research council (NWO) under the talent program (Vidi grant VI.Vidi.203.093).

JRC acknowledges funding from the European Union via the European Research Council (ERC) grant Epaphus (project number: 101166008).

All scripts used to reduce the data, and produce the figures, tables, and results in this work are publicly available on GitHub at \url{https://github.com/ekaterinailin/hip67522-spi/tree/radio-spi/}.
\end{acknowledgements}

\bibliographystyle{aa} 
\bibliography{references.bib} 

\begin{thebibliography}{70}
\expandafter\ifx\csname natexlab\endcsname\relax\def\natexlab#1{#1}\fi

\bibitem[{Barber {et~al.}(2024)Barber, Thao, Mann, Vanderburg, Mori, Livingston, Fukui, Narita, Kraus, Tofflemire, Newton, Winn, Jenkins, Seager, Collins, \& Twicken}]{barber2024tess}
Barber, M.~G., Thao, P.~C., Mann, A.~W., {et~al.} 2024, ApJ, 973, L30

\bibitem[{Barkaoui {et~al.}(2024)Barkaoui, Pozuelos, Hellier, Smalley, Nielsen, Niraula, Gillon, de~Wit, Müller, Dorn, Helled, Jehin, Demory, Van~Grootel, Soubkiou, Ghachoui, Anderson, Benkhaldoun, Bouchy, Burdanov, Delrez, Ducrot, Garcia, Jabiri, Lendl, Maxted, Murray, Pedersen, Queloz, Sebastian, Turner, Udry, Timmermans, Triaud, \& West}]{barkaoui2024extended}
Barkaoui, K., Pozuelos, F.~J., Hellier, C., {et~al.} 2024, NatAs, 8, 909

\bibitem[{Benz {et~al.}(2021)Benz, Broeg, Fortier, Rando, Beck, Beck, Queloz, Ehrenreich, Maxted, Isaak, Billot, Alibert, Alonso, António, Asquier, Bandy, Bárczy, Barrado, Barros, Baumjohann, Bekkelien, Bergomi, Biondi, Bonfils, Borsato, Brandeker, Busch, Cabrera, Cessa, Charnoz, Chazelas, Collier~Cameron, Corral Van~Damme, Cortes, Davies, Deleuil, Deline, Delrez, Demangeon, Demory, Erikson, Farinato, Fossati, Fridlund, Futyan, Gandolfi, Garcia~Munoz, Gillon, Guterman, Gutierrez, Hasiba, Heng, Hernandez, Hoyer, Kiss, Kovacs, Kuntzer, Laskar, Lecavelier~des Etangs, Lendl, López, Lora, Lovis, Lüftinger, Magrin, Malvasio, Marafatto, Michaelis, de~Miguel, Modrego, Munari, Nascimbeni, Olofsson, Ottacher, Ottensamer, Pagano, Palacios, Pallé, Peter, Piazza, Piotto, Pizarro, Pollaco, Ragazzoni, Ratti, Rauer, Ribas, Rieder, Rohlfs, Safa, Salatti, Santos, Scandariato, Ségransan, Simon, Smith, Sordet, Sousa, Steller, Szabó, Szoke, Thomas, Tschentscher, Udry, Van~Grootel, Viotto, Walter, Walton, Wildi, \&
  Wolter}]{benz2021cheops}
Benz, W., Broeg, C., Fortier, A., {et~al.} 2021, Exp. Astron., 51, 109

\bibitem[{Berger {et~al.}(2010)Berger, Basri, Fleming, Giampapa, Gizis, Liebert, Martín, Phan-Bao, \& Rutledge}]{berger2010simultaneous}
Berger, E., Basri, G., Fleming, T.~A., {et~al.} 2010, ApJ, 709, 332

\bibitem[{Blanco-Pozo {et~al.}(2023)Blanco-Pozo, Perger, Damasso, Anglada~Escudé, Ribas, Baroch, Caballero, Cifuentes, Jeffers, Lafarga, Kaminski, Kaur, Nagel, Perdelwitz, Pérez-Torres, Sozzetti, Viganò, Amado, Andreuzzi, Béjar, Brown, Del~Sordo, Dreizler, Galadí-Enríquez, Hatzes, Kürster, Lanza, Melis, Molinari, Montes, Murgia, Pallé, Peña-Moñino, Perrodin, Pilia, Poretti, Quirrenbach, Reiners, Schweitzer, Zapatero~Osorio, \& Zechmeister}]{blanco-pozo2023carmenes}
Blanco-Pozo, J., Perger, M., Damasso, M., {et~al.} 2023, A\&A, 671, A50

\bibitem[{Bloot {et~al.}(2024)Bloot, Callingham, Vedantham, Kavanagh, Pope, Climent, Guirado, Peña-Moñino, \& Pérez-Torres}]{bloot2024phenomenology}
Bloot, S., Callingham, J.~R., Vedantham, H.~K., {et~al.} 2024, A\&A, 682, A170

\bibitem[{Callingham {et~al.}(2024)Callingham, Pope, Kavanagh, Bellotti, Daley-Yates, Damasso, Grießmeier, Güdel, Günther, Kao, Klein, Mahadevan, Morin, Nichols, Osten, Pérez-Torres, Pineda, Rigney, Saur, Stefánsson, Turner, Vedantham, Vidotto, Villadsen, \& Zarka}]{callingham2024radio}
Callingham, J.~R., Pope, B. J.~S., Kavanagh, R.~D., {et~al.} 2024, NatAs, 8, 1359

\bibitem[{Callingham {et~al.}(2021)Callingham, Vedantham, Shimwell, Pope, Davis, Best, Hardcastle, Rottgering, Sabater, Tasse, van Weeren, Williams, Zarka, de~Gasperin, \& Drabent}]{callingham2021population}
Callingham, J.~R., Vedantham, H.~K., Shimwell, T.~W., {et~al.} 2021, NatAs, 5, 1233

\bibitem[{Cauley {et~al.}(2019)Cauley, Shkolnik, Llama, \& Lanza}]{cauley2019magnetic}
Cauley, P.~W., Shkolnik, E.~L., Llama, J., \& Lanza, A.~F. 2019, NatAs, 3, 1128

\bibitem[{Cordun {et~al.}(2025)Cordun, Vedantham, Brentjens, \& van~der Tak}]{cordun2025deep}
Cordun, C.~M., Vedantham, H.~K., Brentjens, M.~A., \& van~der Tak, F. F.~S. 2025, A\&A, 693, A162

\bibitem[{Davis {et~al.}(2024)Davis, Hallinan, Ayala, Dong, \& Myers}]{davis2024detection}
Davis, I., Hallinan, G., Ayala, C., Dong, D., \& Myers, S. 2024, ArXiv e-prints [\eprint{2408.14612}]

\bibitem[{de~Zeeuw {et~al.}(1999)de~Zeeuw, Hoogerwerf, de~Bruijne, Brown, \& Blaauw}]{dezeeuw1999hipparcos}
de~Zeeuw, P.~T., Hoogerwerf, R., de~Bruijne, J. H.~J., Brown, A. G.~A., \& Blaauw, A. 1999, AJ, 117, 354

\bibitem[{Dewdney {et~al.}(2009)Dewdney, Hall, Schilizzi, \& Lazio}]{dewdney2009square}
Dewdney, P.~E., Hall, P.~J., Schilizzi, R.~T., \& Lazio, T. J. L.~W. 2009, IEEE Proceedings, 97, 1482

\bibitem[{Dulk(1985)}]{dulk1985radio}
Dulk, G.~A. 1985, ARA\&A, 23, 169

\bibitem[{Dulk \& Marsh(1982)}]{dulk1982simplified}
Dulk, G.~A. \& Marsh, K.~A. 1982, ApJ, 259, 350

\bibitem[{Ehrenreich {et~al.}(2015)Ehrenreich, Bourrier, Wheatley, Lecavelier~des Etangs, Hébrard, Udry, Bonfils, Delfosse, Désert, Sing, \& Vidal-Madjar}]{ehrenreich2015giant}
Ehrenreich, D., Bourrier, V., Wheatley, P.~J., {et~al.} 2015, Nature, 522, 459

\bibitem[{Fischer \& Saur(2019)}]{fischer2019timevariable}
Fischer, C. \& Saur, J. 2019, ApJ, 872, 113

\bibitem[{{Gaia Collaboration} {et~al.}(2021){Gaia Collaboration}, Brown, Vallenari, Prusti, de~Bruijne, Babusiaux, Biermann, Creevey, Evans, Eyer, Hutton, Jansen, Jordi, Klioner, Lammers, Lindegren, Luri, Mignard, Panem, Pourbaix, Randich, Sartoretti, Soubiran, Walton, Arenou, Bailer-Jones, Bastian, Cropper, Drimmel, Katz, Lattanzi, van Leeuwen, Bakker, Cacciari, Castañeda, De~Angeli, Ducourant, Fabricius, Fouesneau, Frémat, Guerra, Guerrier, Guiraud, Jean-Antoine~Piccolo, Masana, Messineo, Mowlavi, Nicolas, Nienartowicz, Pailler, Panuzzo, Riclet, Roux, Seabroke, Sordo, Tanga, Thévenin, Gracia-Abril, Portell, Teyssier, Altmann, Andrae, Bellas-Velidis, Benson, Berthier, Blomme, Brugaletta, Burgess, Busso, Carry, Cellino, Cheek, Clementini, Damerdji, Davidson, Delchambre, Dell'Oro, Fernández-Hernández, Galluccio, García-Lario, Garcia-Reinaldos, González-Núñez, Gosset, Haigron, Halbwachs, Hambly, Harrison, Hatzidimitriou, Heiter, Hernández, Hestroffer, Hodgkin, Holl, Janßen, Jevardat~de Fombelle,
  Jordan, Krone-Martins, Lanzafame, Löffler, Lorca, Manteiga, Marchal, Marrese, Moitinho, Mora, Muinonen, Osborne, Pancino, Pauwels, Petit, Recio-Blanco, Richards, Riello, Rimoldini, Robin, Roegiers, Rybizki, Sarro, Siopis, Smith, Sozzetti, Ulla, Utrilla, van Leeuwen, van Reeven, Abbas, Abreu~Aramburu, Accart, Aerts, Aguado, Ajaj, Altavilla, Álvarez, Álvarez Cid-Fuentes, Alves, Anderson, Anglada~Varela, Antoja, Audard, Baines, Baker, Balaguer-Núñez, Balbinot, Balog, Barache, Barbato, Barros, Barstow, Bartolomé, Bassilana, Bauchet, Baudesson-Stella, Becciani, Bellazzini, Bernet, Bertone, Bianchi, Blanco-Cuaresma, Boch, Bombrun, Bossini, Bouquillon, Bragaglia, Bramante, Breedt, Bressan, Brouillet, Bucciarelli, Burlacu, Busonero, Butkevich, Buzzi, Caffau, Cancelliere, Cánovas, Cantat-Gaudin, Carballo, Carlucci, Carnerero, Carrasco, Casamiquela, Castellani, Castro-Ginard, Castro~Sampol, Chaoul, Charlot, Chemin, Chiavassa, Cioni, Comoretto, Cooper, Cornez, Cowell, Crifo, Crosta, Crowley, Dafonte,
  Dapergolas, David, David, de~Laverny, De~Luise, De~March, De~Ridder, de~Souza, de~Teodoro, de~Torres, del Peloso, del Pozo, Delbo, Delgado, Delgado, Delisle, Di~Matteo, Diakite, Diener, Distefano, Dolding, Eappachen, Edvardsson, Enke, Esquej, Fabre, Fabrizio, Faigler, Fedorets, Fernique, Fienga, Figueras, Fouron, Fragkoudi, Fraile, Franke, Gai, Garabato, Garcia-Gutierrez, García-Torres, Garofalo, Gavras, Gerlach, Geyer, Giacobbe, Gilmore, Girona, Giuffrida, Gomel, Gomez, Gonzalez-Santamaria, González-Vidal, Granvik, Gutiérrez-Sánchez, Guy, Hauser, Haywood, Helmi, Hidalgo, Hilger, Hładczuk, Hobbs, Holland, Huckle, Jasniewicz, Jonker, Juaristi~Campillo, Julbe, Karbevska, Kervella, Khanna, Kochoska, Kontizas, Kordopatis, Korn, Kostrzewa-Rutkowska, Kruszyńska, Lambert, Lanza, Lasne, Le~Campion, Le~Fustec, Lebreton, Lebzelter, Leccia, Leclerc, Lecoeur-Taibi, Liao, Licata, Lindstrøm, Lister, Livanou, Lobel, Madrero~Pardo, Managau, Mann, Marchant, Marconi, Marcos~Santos, Marinoni, Marocco, Marshall,
  Martin~Polo, Martín-Fleitas, Masip, Massari, Mastrobuono-Battisti, Mazeh, McMillan, Messina, Michalik, Millar, Mints, Molina, Molinaro, Molnár, Montegriffo, Mor, Morbidelli, Morel, Morris, Mulone, Munoz, Muraveva, Murphy, Musella, Noval, Ordénovic, Orrù, Osinde, Pagani, Pagano, Palaversa, Palicio, Panahi, Pawlak, Peñalosa~Esteller, Penttilä, Piersimoni, Pineau, Plachy, Plum, Poggio, Poretti, Poujoulet, Prša, Pulone, Racero, Ragaini, Rainer, Raiteri, Rambaux, Ramos, Ramos-Lerate, Re~Fiorentin, Regibo, Reylé, Ripepi, Riva, Rixon, Robichon, Robin, Roelens, Rohrbasser, Romero-Gómez, Rowell, Royer, Rybicki, Sadowski, Sagristà~Sellés, Sahlmann, Salgado, Salguero, Samaras, Sanchez~Gimenez, Sanna, Santoveña, Sarasso, Schultheis, Sciacca, Segol, Segovia, Ségransan, Semeux, Shahaf, Siddiqui, Siebert, Siltala, Slezak, Smart, Solano, Solitro, Souami, Souchay, Spagna, Spoto, Steele, Steidelmüller, Stephenson, Süveges, Szabados, Szegedi-Elek, Taris, Tauran, Taylor, Teixeira, Thuillot, Tonello, Torra,
  Torra, Turon, Unger, Vaillant, van Dillen, Vanel, Vecchiato, Viala, Vicente, Voutsinas, Weiler, Wevers, Wyrzykowski, Yoldas, Yvard, Zhao, Zorec, Zucker, Zurbach, \& Zwitter}]{gaiacollaboration2021gaiaa}
{Gaia Collaboration}, Brown, A. G.~A., Vallenari, A., {et~al.} 2021, A\&A, 649, A1

\bibitem[{Gordon \& Arnaud(2021)}]{gordon2021pyxspec}
Gordon, C. \& Arnaud, K. 2021, ASCL, ascl:2101.014

\bibitem[{Gudel {et~al.}(1994)Gudel, Schmitt, \& Benz}]{gudel1994discovery}
Gudel, M., Schmitt, J. H. M.~M., \& Benz, A.~O. 1994, Science, 265, 933

\bibitem[{Guedel \& Benz(1993)}]{guedel1993xray}
Guedel, M. \& Benz, A.~O. 1993, ApJ, 405, L63

\bibitem[{Güdel(2004)}]{gudel2004xray}
Güdel, M. 2004, Astron. Astrophys. Rev., 12, 71

\bibitem[{Hallinan {et~al.}(2008)Hallinan, Antonova, Doyle, Bourke, Lane, \& Golden}]{hallinan2008confirmation}
Hallinan, G., Antonova, A., Doyle, J.~G., {et~al.} 2008, ApJ, 684, 644

\bibitem[{Hallinan {et~al.}(2015)Hallinan, Littlefair, Cotter, Bourke, Harding, Pineda, Butler, Golden, Basri, Doyle, Kao, Berdyugina, Kuznetsov, Rupen, \& Antonova}]{hallinan2015magnetospherically}
Hallinan, G., Littlefair, S.~P., Cotter, G., {et~al.} 2015, Nature, 523, 568

\bibitem[{Heitzmann {et~al.}(2021)Heitzmann, Zhou, Quinn, Marsden, Wright, Petit, Vanderburg, Bouma, Mann, \& Rizzuto}]{heitzmann2021obliquity}
Heitzmann, A., Zhou, G., Quinn, S.~N., {et~al.} 2021, ApJL, 922, L1

\bibitem[{Hess \& Zarka(2011)}]{hess2011modeling}
Hess, S. L.~G. \& Zarka, P. 2011, A\&A, 531, A29

\bibitem[{Ilin \& Poppenhaeger(2022)}]{ilin2022searching}
Ilin, E. \& Poppenhaeger, K. 2022, MNRAS, 513, 4579

\bibitem[{Ilin {et~al.}(2024)Ilin, Poppenhäger, Chebly, Ilić, \& Alvarado-Gómez}]{ilin2024planetary}
Ilin, E., Poppenhäger, K., Chebly, J., Ilić, N., \& Alvarado-Gómez, J.~D. 2024, MNRAS, 527, 3395

\bibitem[{Ilin {et~al.}(2025)Ilin, Vedantham, Poppenhaeger, Bloot, Callingham, Brandeker, \& Chakraborty}]{ilin2025closein}
Ilin, E., Vedantham, H., Poppenhaeger, K., {et~al.} 2025, Nature, in press.

\bibitem[{Kavanagh {et~al.}(2021)Kavanagh, Vidotto, Klein, Jardine, Donati, \& Ó~Fionnagáin}]{kavanagh2021planetinduced}
Kavanagh, R.~D., Vidotto, A.~A., Klein, B., {et~al.} 2021, MNRAS, 504, 1511

\bibitem[{Ketzer \& Poppenhäger(2023)}]{ketzer2023influence}
Ketzer, L. \& Poppenhäger, K. 2023, MNRAS, 518, 1683

\bibitem[{Kochukhov {et~al.}(2020)Kochukhov, Hackman, Lehtinen, \& Wehrhahn}]{kochukhov2020hidden}
Kochukhov, O., Hackman, T., Lehtinen, J.~J., \& Wehrhahn, A. 2020, A\&A, 635, A142

\bibitem[{Lanza(2018)}]{lanza2018closeby}
Lanza, A.~F. 2018, A\&A, 610, A81

\bibitem[{Laughlin {et~al.}(2011)Laughlin, Crismani, \& Adams}]{laughlin2011anomalous}
Laughlin, G., Crismani, M., \& Adams, F.~C. 2011, ApJ, 729, L7

\bibitem[{Lecavelier Des~Etangs {et~al.}(2010)Lecavelier Des~Etangs, Ehrenreich, Vidal-Madjar, Ballester, Désert, Ferlet, Hébrard, Sing, Tchakoumegni, \& Udry}]{lecavelierdesetangs2010evaporation}
Lecavelier Des~Etangs, A., Ehrenreich, D., Vidal-Madjar, A., {et~al.} 2010, A\&A, 514, A72

\bibitem[{Maggio {et~al.}(2024)Maggio, Pillitteri, Argiroffi, Locci, Benatti, \& Micela}]{maggio2024xuv}
Maggio, A., Pillitteri, I., Argiroffi, C., {et~al.} 2024, A\&A, 690, A383

\bibitem[{Martioli {et~al.}(2021)Martioli, Hébrard, Correia, Laskar, \& Lecavelier~des Etangs}]{martioli2021new}
Martioli, E., Hébrard, G., Correia, A. C.~M., Laskar, J., \& Lecavelier~des Etangs, A. 2021, A\&A, 649, A177

\bibitem[{Melrose \& Dulk(1982)}]{melrose1982electroncyclotron}
Melrose, D.~B. \& Dulk, G.~A. 1982, ApJ, 259, 844

\bibitem[{Offringa {et~al.}(2014)Offringa, McKinley, Hurley-Walker, Briggs, Wayth, Kaplan, Bell, Feng, Neben, Hughes, Rhee, Murphy, Bhat, Bernardi, Bowman, Cappallo, Corey, Deshpande, Emrich, Ewall-Wice, Gaensler, Goeke, Greenhill, Hazelton, Hindson, Johnston-Hollitt, Jacobs, Kasper, Kratzenberg, Lenc, Lonsdale, Lynch, McWhirter, Mitchell, Morales, Morgan, Kudryavtseva, Oberoi, Ord, Pindor, Procopio, Prabu, Riding, Roshi, Shankar, Srivani, Subrahmanyan, Tingay, Waterson, Webster, Whitney, Williams, \& Williams}]{offringa2014wsclean}
Offringa, A.~R., McKinley, B., Hurley-Walker, N., {et~al.} 2014, MNRAS, 444, 606

\bibitem[{Osten {et~al.}(2005)Osten, Hawley, Allred, Johns-Krull, \& Roark}]{osten2005radio}
Osten, R.~A., Hawley, S.~L., Allred, J.~C., Johns-Krull, C.~M., \& Roark, C. 2005, ApJ, 621, 398

\bibitem[{Perley {et~al.}(2011)Perley, Chandler, Butler, \& Wrobel}]{perley2011expanded}
Perley, R.~A., Chandler, C.~J., Butler, B.~J., \& Wrobel, J.~M. 2011, ApJL, 739, L1

\bibitem[{Pineda \& Villadsen(2023)}]{pineda2023coherent}
Pineda, J.~S. \& Villadsen, J. 2023, NatAs, 7, 569

\bibitem[{Plavchan {et~al.}(2020)Plavchan, Barclay, Gagné, Gao, Cale, Matzko, Dragomir, Quinn, Feliz, Stassun, Crossfield, Berardo, Latham, Tieu, Anglada-Escudé, Ricker, Vanderspek, Seager, Winn, Jenkins, Rinehart, Krishnamurthy, Dynes, Doty, Adams, Afanasev, Beichman, Bottom, Bowler, Brinkworth, Brown, Cancino, Ciardi, Clampin, Clark, Collins, Davison, Foreman-Mackey, Furlan, Gaidos, Geneser, Giddens, Gilbert, Hall, Hellier, Henry, Horner, Howard, Huang, Huber, Kane, Kenworthy, Kielkopf, Kipping, Klenke, Kruse, Latouf, Lowrance, Mennesson, Mengel, Mills, Morton, Narita, Newton, Nishimoto, Okumura, Palle, Pepper, Quintana, Roberge, Roccatagliata, Schlieder, Tanner, Teske, Tinney, Vanderburg, von Braun, Walp, Wang, Wang, Weigand, White, Wittenmyer, Wright, Youngblood, Zhang, \& Zilberman}]{plavchan2020planet}
Plavchan, P., Barclay, T., Gagné, J., {et~al.} 2020, Nature, 582, 497

\bibitem[{Pope {et~al.}(2021)Pope, Callingham, Feinstein, Günther, Vedantham, Ansdell, \& Shimwell}]{pope2021tess}
Pope, B. J.~S., Callingham, J.~R., Feinstein, A.~D., {et~al.} 2021, ApJ, 919, L10

\bibitem[{Poppenhaeger \& Schmitt(2011)}]{poppenhaeger2011correlation}
Poppenhaeger, K. \& Schmitt, J. H. M.~M. 2011, ApJ, 735, 59

\bibitem[{Pérez-Torres {et~al.}(2021)Pérez-Torres, Gómez, Ortiz, Leto, Anglada, Gómez, Rodríguez, Trigilio, Amado, Alberdi, Anglada-Escudé, Osorio, Umana, Berdiñas, López-González, Morales, Rodríguez-López, \& Chibueze}]{perez-torres2021monitoring}
Pérez-Torres, M., Gómez, J.~F., Ortiz, J.~L., {et~al.} 2021, A\&A, 645, A77

\bibitem[{Reiners {et~al.}(2022)Reiners, Shulyak, Käpylä, Ribas, Nagel, Zechmeister, Caballero, Shan, Fuhrmeister, Quirrenbach, Amado, Montes, Jeffers, Azzaro, Béjar, Chaturvedi, Henning, Kürster, \& Pallé}]{reiners2022magnetism}
Reiners, A., Shulyak, D., Käpylä, P.~J., {et~al.} 2022, A\&A, 662, A41

\bibitem[{Ricker {et~al.}(2015)Ricker, Winn, Vanderspek, Latham, Bakos, Bean, Berta-Thompson, Brown, Buchhave, Butler, Butler, Chaplin, Charbonneau, Christensen-Dalsgaard, Clampin, Deming, Doty, De~Lee, Dressing, Dunham, Endl, Fressin, Ge, Henning, Holman, Howard, Ida, Jenkins, Jernigan, Johnson, Kaltenegger, Kawai, Kjeldsen, Laughlin, Levine, Lin, Lissauer, MacQueen, Marcy, McCullough, Morton, Narita, Paegert, Palle, Pepe, Pepper, Quirrenbach, Rinehart, Sasselov, Sato, Seager, Sozzetti, Stassun, Sullivan, Szentgyorgyi, Torres, Udry, \& Villasenor}]{ricker2015transiting}
Ricker, G.~R., Winn, J.~N., Vanderspek, R., {et~al.} 2015, J. Astron. Telesc. Instrum. Syst., 1, 014003

\bibitem[{Rizzuto {et~al.}(2020)Rizzuto, Newton, Mann, Tofflemire, Vanderburg, Kraus, Wood, Quinn, Zhou, Thao, Law, Ziegler, \& Briceño}]{rizzuto2020tess}
Rizzuto, A.~C., Newton, E.~R., Mann, A.~W., {et~al.} 2020, AJ, 160, 33

\bibitem[{Saur {et~al.}(2013)Saur, Grambusch, Duling, Neubauer, \& Simon}]{saur2013magnetic}
Saur, J., Grambusch, T., Duling, S., Neubauer, F.~M., \& Simon, S. 2013, A\&A, 552, A119

\bibitem[{Scharf(2010)}]{scharf2010possible}
Scharf, C.~A. 2010, ApJ, 722, 1547

\bibitem[{Shkolnik {et~al.}(2008)Shkolnik, Bohlender, Walker, \& Collier~Cameron}]{shkolnik2008nature}
Shkolnik, E., Bohlender, D.~A., Walker, G. A.~H., \& Collier~Cameron, A. 2008, ApJ, 676, 628

\bibitem[{Shkolnik \& Llama(2018)}]{shkolnik2018signatures}
Shkolnik, E.~L. \& Llama, J. 2018, Signatures of {Star}-{Planet} {Interactions}, Handbook of {Exoplanets} (Springer International Publishing AG)

\bibitem[{Spake {et~al.}(2021)Spake, Oklopčić, \& Hillenbrand}]{spake2021posttransit}
Spake, J.~J., Oklopčić, A., \& Hillenbrand, L.~A. 2021, AJ, 162, 284

\bibitem[{Strugarek {et~al.}(2022)Strugarek, Fares, Bourrier, Brun, Réville, Amari, Helling, Jardine, Llama, Moutou, Vidotto, Wheatley, \& Zarka}]{strugarek2022moves}
Strugarek, A., Fares, R., Bourrier, V., {et~al.} 2022, MNRAS, 512, 4556

\bibitem[{Team {et~al.}(2022)Team, Bean, Bhatnagar, Castro, Meyer, Emonts, Garcia, Garwood, Golap, Villalba, Harris, Hayashi, Hoskins, Hsieh, Jagannathan, Kawasaki, Keimpema, Kettenis, Lopez, Marvil, Masters, McNichols, Mehringer, Miel, Moellenbrock, Montesino, Nakazato, Ott, Petry, Pokorny, Raba, Rau, Schiebel, Schweighart, Sekhar, Shimada, Small, Steeb, Sugimoto, Suoranta, Tsutsumi, Bemmel, Verkouter, Wells, Xiong, Szomoru, Griffith, Glendenning, \& Kern}]{team2022casa}
Team, T.~C., Bean, B., Bhatnagar, S., {et~al.} 2022, PASP, 134, 114501

\bibitem[{Thao {et~al.}(2024)Thao, Mann, Feinstein, Gao, Thorngren, Rotman, Welbanks, Brown, Duvvuri, France, Longo, Sandoval, Schneider, Wilson, Youngblood, Vanderburg, Barber, Wood, Batalha, Kraus, Murray, Newton, Rizzuto, Tofflemire, Tsai, Bean, Berta-Thompson, Evans-Soma, Froning, Kempton, Miguel, \& Pineda}]{thao2024featherweight}
Thao, P.~C., Mann, A.~W., Feinstein, A.~D., {et~al.} 2024, AJ, 168, 297

\bibitem[{Tian(2015)}]{tian2015atmospheric}
Tian, F. 2015, Annu. Rev. Earth Planet. Sci., 43, 459

\bibitem[{Treumann(2006)}]{treumann2006electroncyclotron}
Treumann, R.~A. 2006, Astron. Astrophys. Rev., 13, 229

\bibitem[{Truemper(1982)}]{truemper1982rosat}
Truemper, J. 1982, Advances in Space Research, 2, 241

\bibitem[{Turner {et~al.}(2024)Turner, Grießmeier, Zarka, Zhang, \& Mauduit}]{turner2024followup}
Turner, J.~D., Grießmeier, J.-M., Zarka, P., Zhang, X., \& Mauduit, E. 2024, A\&A, 688, A66

\bibitem[{Turner {et~al.}(2021)Turner, Zarka, Grießmeier, Lazio, Cecconi, Emilio~Enriquez, Girard, Jayawardhana, Lamy, Nichols, \& de~Pater}]{turner2021search}
Turner, J.~D., Zarka, P., Grießmeier, J.-M., {et~al.} 2021, A\&A, 645, A59

\bibitem[{van Diepen {et~al.}(2018)van Diepen, Dijkema, \& Offringa}]{vandiepen2018dppp}
van Diepen, G., Dijkema, T.~J., \& Offringa, A. 2018, ASCL, ascl:1804.003

\bibitem[{Van~Looveren {et~al.}(2024)Van~Looveren, Güdel, Boro~Saikia, \& Kislyakova}]{vanlooveren2024airy}
Van~Looveren, G., Güdel, M., Boro~Saikia, S., \& Kislyakova, K. 2024, A\&A, 683, A153

\bibitem[{Vedantham {et~al.}(2020)Vedantham, Callingham, Shimwell, Tasse, Pope, Bedell, Snellen, Best, Hardcastle, Haverkorn, Mechev, O’Sullivan, Röttgering, \& White}]{vedantham2020coherent}
Vedantham, H.~K., Callingham, J.~R., Shimwell, T.~W., {et~al.} 2020, NatAs, 4, 577

\bibitem[{Williams {et~al.}(2014)Williams, Cook, \& Berger}]{williams2014trends}
Williams, P. K.~G., Cook, B.~A., \& Berger, E. 2014, ApJ, 785, 9

\bibitem[{Wilson {et~al.}(2011)Wilson, Ferris, Axtens, Brown, Davis, Hampson, Leach, Roberts, Saunders, Koribalski, Caswell, Lenc, Stevens, Voronkov, Wieringa, Brooks, Edwards, Ekers, Emonts, Hindson, Johnston, Maddison, Mahony, Malu, Massardi, Mao, McConnell, Norris, Schnitzeler, Subrahmanyan, Urquhart, Thompson, \& Wark}]{wilson2011australia}
Wilson, W.~E., Ferris, R.~H., Axtens, P., {et~al.} 2011, MNRAS, 416, 832

\bibitem[{Yiu {et~al.}(2024)Yiu, Vedantham, Callingham, \& Günther}]{yiu2024radio}
Yiu, T. W.~H., Vedantham, H.~K., Callingham, J.~R., \& Günther, M.~N. 2024, A\&A, 684, A3

\bibitem[{Zarka(2007)}]{zarka2007plasma}
Zarka, P. 2007, P\&SS, 55, 598

\bibitem[{Zarka {et~al.}(2001)Zarka, Treumann, Ryabov, \& Ryabov}]{zarka2001magneticallydriven}
Zarka, P., Treumann, R.~A., Ryabov, B.~P., \& Ryabov, V.~B. 2001, Astrophys. Space Sci., 277, 293

\end{thebibliography}

\begin{appendix}
\section{Stokes I detection}
    \begin{figure}[h]
        \centering
        \includegraphics[width=0.9\linewidth]{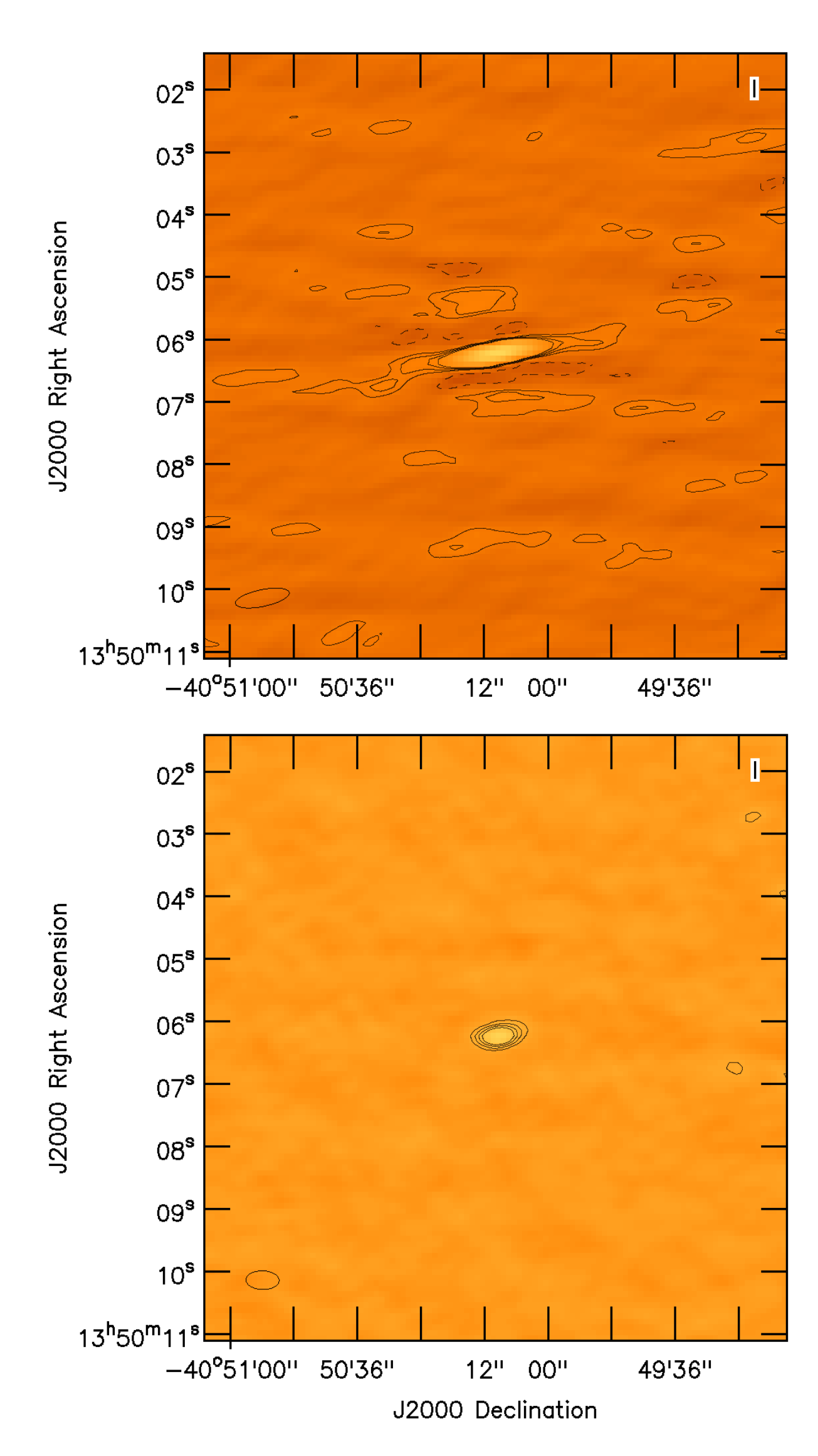}
        \caption{HIP 67522 detected in Stokes I with ATCA during the June 11 burst (top panel), and in quiescence on June 19 (bottom panel). Bottom left oval shows the beam in each panel. Contours represent the 3, 5, 7, and 9 RMS noise levels.}
        \label{fig:Appendix}
    \end{figure}
\end{appendix}

\end{document}